Research article

Jingyi Yang, Indra Ghimire, Pin Chieh Wu, Sudip Gurung, Catherine Arndt, Din Ping Tsai and Ho Wai Howard Lee*

# Photonic crystal fiber metalens

**Abstract:** Conventional optical fiber has excellent performance in guiding light, which has been widely employed for long-distance optical communication. Although the optical fiber is efficient for transmitting light, its functionality is limited by the dielectric properties of the core's and cladding's materials (e.g. Ge-doped-silica and silica glasses). The spot size of the transmitted light is diverging and restricted by the diffraction limit of the dielectric core, and the numerical aperture is determined by the refractive index of the fiber materials. However, the novel technology of metasurfaces is opening the door to a variety of optical fiber innovations. Here, we report an ultrathin optical metalens directly patterned on the facet of a photonic crystal optical fiber that enables light focusing in the telecommunication regime. In-fiber metalenses with focal lengths of 28 µm and 40 µm at a wavelength of 1550 nm are demonstrated with maximum enhanced optical intensity as large as 234%. The ultrathin optical fiber metalens may find novel applications in optical imaging, sensing, and fiber laser designs.

*\*Corresponding author: Ho Wai Howard Lee,* Department of Physics and Baylor Research and Innovation Collaborative (BRIC), Baylor University, Waco, TX 76798, USA; and The Institute for Quantum Science and Engineering, Texas A&M University, College Station, TX 77843, USA, e-mail: Howard_Lee@baylor.edu
**Jingyi Yang, Indra Ghimire, Sudip Gurung and Catherine Arndt:** Department of Physics and Baylor Research and Innovation Collaborative (BRIC), Baylor University, Waco, TX 76798, USA. https://orcid.org/0000-0001-8219-9460 (J. Yang)
**Pin Chieh Wu:** Research Center for Applied Sciences, Academia Sinica, Taipei 11529, Taiwan; and Thomas J. Watson Laboratories of Applied Physics, California Institute of Technology, Pasadena, CA 91125, USA
**Din Ping Tsai:** Research Center for Applied Sciences, Academia Sinica, Taipei 11529, Taiwan

## 1 Introduction

Optical fiber is a well-established efficient platform used to guide light and allow high bandwidth optical transmission for long-distance communication with low attenuation. Besides, conventional optical fibers also have been widely employed for fiber lasers [1, 2], remote and optical sensing [3], fiber imaging [4] and endoscope [5], and fiber laser surgery [6, 7]. However, the optical properties of fiber waveguide, such as phase, amplitude, polarization state, and mode profile, cannot be altered after the fiber drawing fabrication. In addition, the spot size of the transmitted light is divergent and restricted by the diffraction limit of the dielectric core. Attempts have been made to fabricate periodical plasmonic nanostructures (i.e. slits, holes, bars, etc.) on the optical fiber facets to alter the optical properties and to extend the functionalities of the fibers, as elements of these plasmonic nanostructures can interact directly with well-guided modes of the optical fiber. Compact optical fiber components such as diffraction grating [8, 9], amplifier [10], nanotrimmer [11], optical tweezer [12], and plasmonic sensors [13–15] have been realized with periodical nanostructures on facets of the optical fibers. A method to apply a metallic structure to a polymeric membrane on the facet of a hollow core optical fiber has been functionalized as a nanoplasmonic filter [16]. Realization of an in-fiber focusing lens via plasmonic nanostructures has also been studied in recent years [17–19]. However, these in-fiber plasmonic lenses with concentric annular slits suffer from relatively short focal length (<10 µm), narrow operation bandwidth, and side band caused by high-order diffraction, thus limiting their potential practical applications.

The emergence of metasurfaces provides the opportunity to tailor optical properties for advanced light manipulation and to develop novel ultrathin optical devices [20–22]. By producing a specific phase profile using spatially varied nanoantenna elements, metasurfaces can control the wavefront of the transmitted, reflected, scattered light and enable novel ultrathin optical components such as flat metalenses [23–26]. With the ability to control the optical phase via metasurfaces, several in-fiber metasurface optical components [27] have been demonstrated, for instance, an in-fiber modulator [28]/linear polarizer [29] and an in-fiber beam element [30]. However, to date, no direct integration of metasurface-based lens onto the optical fiber has been demonstrated without relying on a prism or flat substrates. In this work, we developed the first optical fiber metalens with engineerable focal length by directly patterning a geometric phase (i.e. Berry phase) altering metasurface onto the end facet of a large-mode-area photonic crystal fiber (LMA-PCF). To integrate the sufficient area and phase profile of the metalens, instead of using conventional single-mode step index fiber, a LMA-PCF was employed as the platform to guide the light in large core diameter while maintaining the single mode property (endlessly single mode) via the modified total internal reflection mechanism. We experimentally demonstrated that the circularly polarized incident beam can be focused after exiting the fiber in the telecommunication wavelength regime with considerable focusing efficiency. Our work shows a first proof-of-principle demonstration for developing an efficient in-fiber metalens, which will find practical applications for in-fiber optical devices with unique functionalities.

## 2 Design and fabrication

The required distribution of phase retardation for a metalens typically follows:

$$\varphi(r,\lambda) = -\frac{2\pi}{\lambda}(\sqrt{r^2+f^2}-f), \quad (1)$$

where $r$ is the distance from current location on the metalens to the center, $f$ is the focal length, and $\lambda$ is the operation wavelength. With such a phase profile, the incident plane wavefronts are transformed into spherical ones, which converge at the focal length. At a fixed focal length $f$ and an operation wavelength $\lambda$, the phase modulation $\varphi(r,\lambda)$ can be achieved by geometric phase method of the resonant elements in the metasurfaces (Figure 1). The circularly polarized light transmitted through a unit element with the rotational angle $\theta$ acquires an additional phase $\pm 2\theta$, i.e. geometric phase, where the sign indicates the handedness of circularly polarized incidence [31]. Our metalens is designed on LMA-PCF (LMA-25), which is made of pure silica with a core diameter of $25\pm1$ μm (Figure 1A and B). Compared to conventional single-mode fiber (SMF-28) with the core diameter of 8.2 μm, the larger core area of PCF allows more unit elements (total 1261) to be fabricated on the core, thus providing the essential

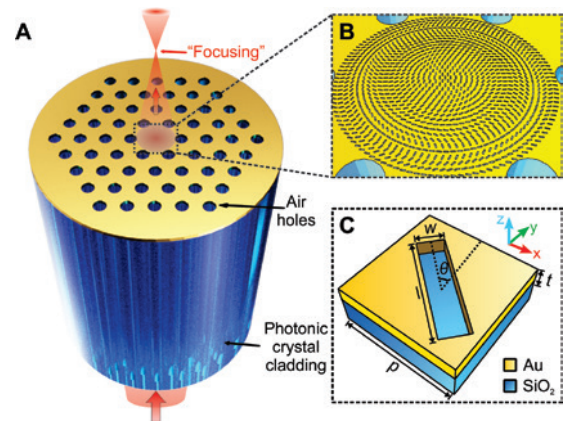

**Figure 1:** Photonic crystal fiber metalens.
(A, B) Schematics of in-fiber metalens on LMA-PCF and (C) unit element of PCF metalens.

phase profile for achieving focusing functionality than conventional single-mode fiber.

The simulated fundamental mode profile (i.e. electric field intensity distribution) of the LMA-PCF is shown in Figure 2A (see *Methods* for more properties of the LMA-PCF). Endlessly, single-mode guiding is maintained in the LMA-PCF with a broad wavelength range, and the light is confined mostly within the core region [32, 33]. The designed metalens is the same size as the core to ensure that the guided core mode interacts with the entire metasurface. With the dimension of the LMA-PCF, we performed theoretical simulations to determine the required rotational angle of each individual resonant element with respect to the center to achieve the convergent effect for certain circularly polarized incident light. A single etched gold nanorod was considered as a unit element. The simulated conversion efficiency of the optimized unit element (with length of 500 nm, width of 150 nm, periodicity of 600 nm, thickness of 40 nm) for constructing the metalenses is shown in Figure 2B. Its resonant wavelength is located at 1482.6 nm with maximum efficiency of 17%. The electrical field intensity distribution at resonant wavelength is shown in the inset of Figure 2B. The rotational angles $\theta$ of the unit elements as a function of their radial positions that are required for constructing the metalenses are shown in Figure 2C. The designed focal lengths are 30 μm and 50 μm with numerical apertures (NAs) of 0.37 and 0.23, respectively, at an operating wavelength of 1550 nm (see Supplementary S1 for the required rotational angles for two PCF metalenses and *Methods* for details of the simulation).

To experimentally realize the PCF metalens, we first deposited a gold layer with a thickness of 40 nm on the end facet of the LMA-PCF by magneton sputtering. The metalens pattern was fabricated by focused ion beam (FIB) milling with an acceleration voltage of 30 kV and current of 1.5 pA. Special care was taken to align the center of the fiber such that the metasurface pattern completely covered the core of the fiber. The scanning electron microscope (SEM) images of the fabricated PCF metalens are depicted in Figure 3 (Figure 3A and B: NA=0.37, Figure 3C and D: NA=0.23). The simulated transmissions of the fabricated samples using the dimensions obtained in SEM show that the resonant wavelengths are located at 1499.3 nm and 1490.3 nm with maximum efficiencies of 16.9% and 16.5% for metalens with NAs of 0.37 and 0.23, respectively. (See Supplementary S1 for details of the experimental structural parameters, the simulated transmission of the unit cells, and the electric field intensity distributions of the corresponding elements.)

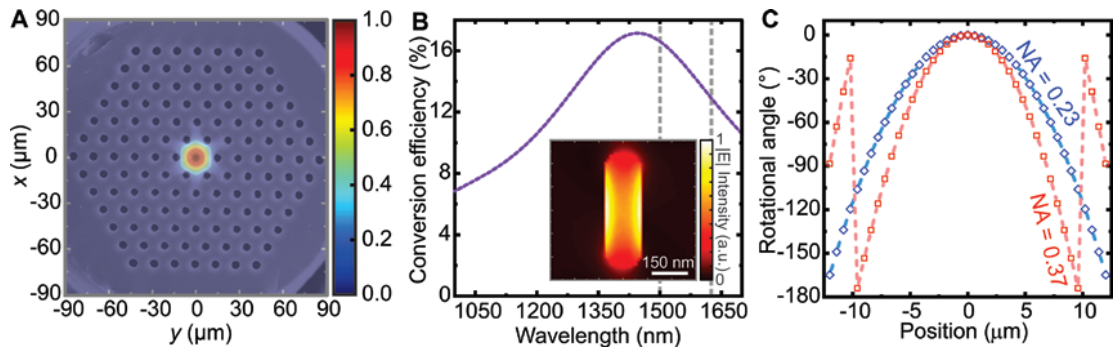

**Figure 2:** Characterization of optical fiber and designed metasurfaces.
(A) Simulated fundamental mode profile (i.e. electric field intensity distribution) superimposed with a SEM image of the LMA-PCF. (B) RCP to LCP efficiency of optimized unit element ($l=500$ nm, $w=150$ nm, $p=600$ nm, $t=40$ nm) in the PCF metalens in simulation. The wavelength region highlighted between gray dash lines marks the experimental bandwidth. Inset: electric field intensity distribution of the PCF metalens' optimized unit elements at corresponding resonance wavelength of 1482.6 nm. (C) Required rotational angle $\theta$ of unit element along the fiber core for PCF metalens for NA of 0.37 (designed focal length of 30 μm) and 0.23 (designed focal length of 50 μm) at a wavelength of 1550 nm.

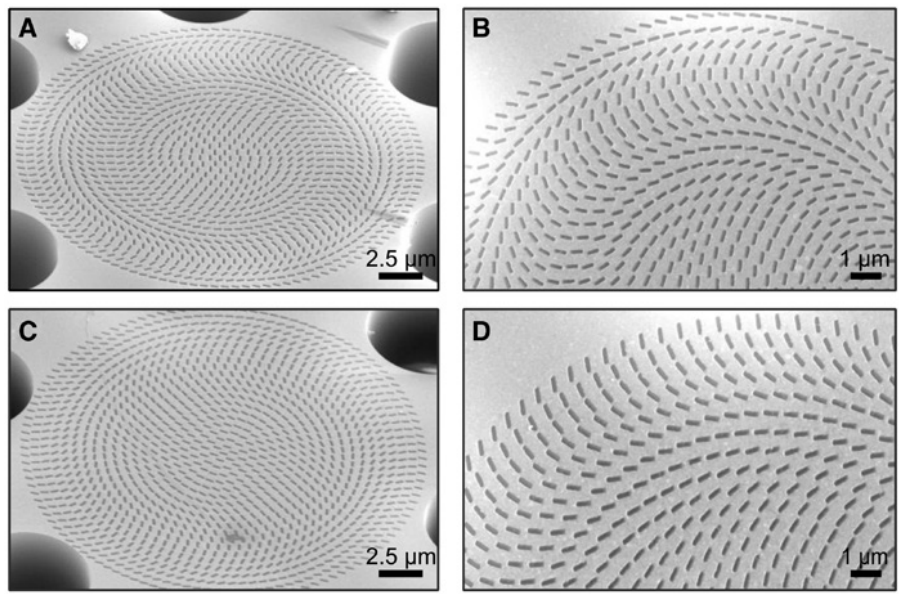

**Figure 3:** SEM images of fabricated PCF metalens for NAs of (A, B) 0.37 and (C, D) 0.23.
The core diameter of the LMA-PCF is 25 ± 1 μm, providing the area for sufficient metasurface unit elements to ensure smooth phase distribution (see Supplementary Table S1 for parameters of the structural parameters of the unit elements for the two LMA-PCF metalenses).

## 3 Results and discussion

To verify the focusing effect of the in-fiber metalens, we captured the light intensity distributions by imaging the mode intensity along the light propagation direction with a z-scan setup that consists of a quarter waveplate, linear polarizer, high NA objective, and near-infrared (NIR) camera (see Supplementary S2 for details of the experimental setup). Using the designed geometric-phase-based metalenses, the incident right-hand circular polarized (RCP) light was launched into the PCF metalens, and the left-hand circular polarization (LCP) component was collected as output light. The stitching mode intensity profiles on the x–z plane for these two PCF metalenses at wavelengths from 1500 nm to 1600 nm with a step of 50 nm are shown in Figure 4A and C. It can be seen that the light density increases as the distance increases from the end facet of the PCF metalens to the focal plane, thus demonstrating the focusing effect. The observed focal lengths are 30 μm and 40 μm, respectively. The measured light distributions show good agreement with the output intensities of the LCP components obtained from the simulation (Figure 4B and D). The actual size parameters of the unit elements obtained from the SEM images of two PCF metalenses are utilized in the simulation (see Supplementary S3 for more experimental and simulated light intensity distributions of PCF metalenses with NAs of 0.37 and 0.23 from an operation wavelength of 1500 nm to 1630 nm with a step of 10 nm). To confirm the focusing effect from the metasurfaces, the RCP components from the output were collected in the same way as LCP components. We extracted mode intensity distributions along the core diameter from NIR camera images for two PCF metalenses correspondingly. The mode intensities were fitted by Gaussian function and normalized to the maximum mode intensity of RCP-LCP at corresponding operation wavelengths. The mode intensity distributions of both the RCP-LCP and RCP-RCP input/output combinations at corresponding focal planes for two metalenses at 1500 nm, 1550 nm, and 1600 nm were recorded and are depicted in Figure 4E and F, respectively. It is clear from the figure that a focusing effect of the metalens is observed only for the input/output combination of RCP-LCP. The slight amount of light measured in the RCP-RCP combination is attributed to the imperfect filtering/alignment of the input and output polarizers and the unmaintained RCP light within the PCF (see Supplementary S4 for more experimental mode intensity distributions of PCF metalenses with NAs of 0.37 and 0.23 for operation wavelength from 1500 nm to 1630 nm with a step of 10 nm). No focusing was observed from the RCP output component, which is consistent with the theoretical design of the Berry-phase-based metalens [23, 24, 34, 35].

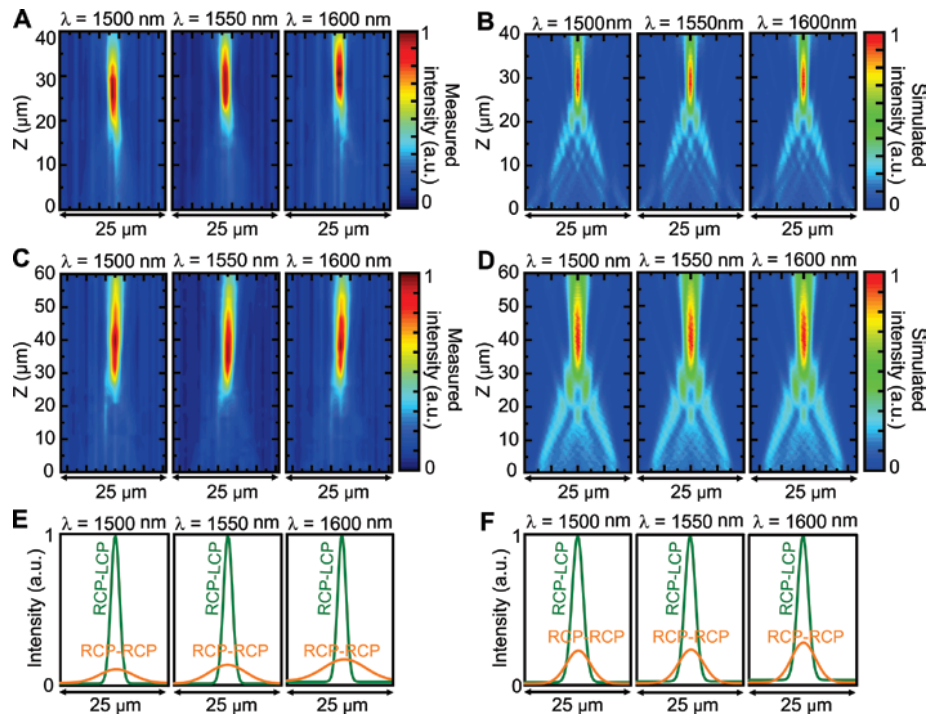

**Figure 4:** Intensity distributions of PCF metalenses.
Measured intensity and simulated intensity profiles of PCF metalenses with (A, B) NA = 0.37 and (C, D) NA = 0.23 along the light propagation direction at various incident wavelengths. Experimental intensity distributions of each mode at corresponding focal planes for RCP-LCP and RCP-RCP input/output combinations of the PCF metalens with NA of (E) 0.37 and (F) 0.23 at different wavelengths.

We further analyzed the optical performance of metalenses. Figure 5A shows the analysis of the dependence of focal lengths and full-width at half-maximum (FWHM) of the focal spot on the operation wavelength. For the PCF metalens with NA of 0.37, the measured focal lengths varied from 26.7 μm to 28.0 μm between wavelengths of 1500 and 1630 nm, which are in good agreement with the simulated results (i.e. focal length of 28.5 μm within the same wavelength range). For the metalens with NA of 0.23, the experimental focal lengths are varied from 40.0 μm to 41.3 μm between operation wavelengths of 1500 and 1630 nm, closely resembling the simulated results of 40.0 μm to 41.0 μm. However, this focal length shows larger offset from the theoretical calculation (i.e. the designed focal length, 50 μm), which is mainly from the implemented phase distribution only divided in the first Fresnel zone. This can be effectively addressed by either increasing the core size or the NA of metalens. The measured and simulated focal lengths at the wavelength of 1550 nm are 28.0 μm and 40.0 μm for the two PCF metalenses, which are close to our simulation. The FWHM is defined as the beam waist of half-maximum light intensity, which is obtained by fitting the measured cross-section of light intensity at the focal spot with the Gaussian function. The measured FWHM values are 2.40–2.63 μm and 3.44–3.65 μm in the measured wavelength range for the PCF metalenses with NAs of 0.37 and 0.23, respectively (see Supplementary S5 for a comparison of spot sizes between two PCF metalenses and LMA-PCF without gold metalens).

The operating efficiency and enhanced optical intensity of the PCF metalenses are shown in Figure 5B. Operating efficiency (i.e. the focusing efficiency from incident RCP to output LCP) is defined as the ratio of the light intensity integrated over the whole beam spot at the focal plane of PCF metalenses to that at the endface of a reference PCF without gold coating. The maximum efficiencies of 16.1% and 16.4% were measured for these two PCF metalenses at wavelengths of 1530 nm and 1540 nm, respectively. The decrease in the operation efficiency with wavelength is in accordance with simulated RCP-to-LCP conversion efficiency from the unit element (Figure 2B). It should be noted that the operation efficiency can be further enhanced by optimizing the configuration or employing dielectric metasurfaces [20, 36]. Operation efficiency as high as 91% could be obtained with optimized parameters of Si nanoantennas (see Supplementary S6 for details). Those Si nano-antennas could be implemented onto the fiber with similar nanofabrication techniques such as FIB milling and electron beam lithography [11].

We also defined the enhanced optical intensity of the PCF metalenses as the ratio of averaged light intensity over the whole beam spot at the focal plane to that at the end-facet of the LMA-PCF without metalens (Figure 5B). The enhanced optical intensity is 234% larger at the wavelength of 1540 nm. Even though the focusing performance is limited by operation efficiency, the light intensity of the PCF metalens is more intense than that of the reference fiber. The demonstrated efficient in-fiber metalens provides an engineerable NA and FWHM with enhanced optical intensity. It could enhance the development of novel and ultracompact in-fiber optical imaging, sensing, and communication device applications such as in-fiber lenses [37, 38] for laser surgery and optical fiber endoscopes [26], a focusing element for optical fiber laser and spectroscopy, and an efficient fiber coupling for optical interconnects. The optical metalens could further be extended to hollow core optical fibers by integrating with a flexible membrane on the hollow core [16], providing a unique focusing element for a fiber gas laser. Further integration of electrically tunable materials such as transparent conducting oxide materials into the in-fiber metalens [39, 40] could potentially enable tuning of the focusing effect or focal length, which would be significant for novel fiber optical trapping and active sensing.

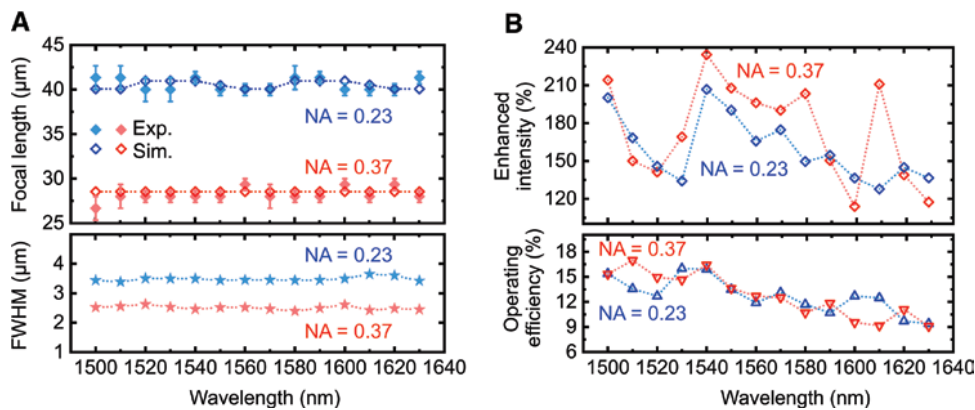

**Figure 5:** Characteristics of PCF metalenses.
(A) Measured, simulated focal lengths and FWHM and (B) operating efficiency and enhanced optical intensity of PCF metalenses with NA = 0.23 and NA = 0.37 at corresponding focal plane versus wavelength.

## 4 Conclusions

We experimentally demonstrated an in-fiber metalens enabled by ultrathin geometric phase metasurfaces with thickness of 40 nm fabricated on the end-facet of the photonic crystal fiber. The ultrathin metalens provides a phase gradient and focuses light from the fiber as a refractive converging lens at telecommunication wavelengths. The focal lengths 28 μm and 40 μm at the operating wavelength of 1550 nm were measured, providing light focusing with NAs of 0.37 and 0.23, respectively. The maximum operating efficiency is 16.4%, which approaches the theoretically predicted level for flat metallic metasurfaces. The maximum enhanced optical intensity is as large as 234%. This integration of metalens and optical fiber will pave the way for in-fiber optical imaging and sensing applications and be significant in the miniaturizing of optical fiber devices with advanced multifunctionalities.

## 5 Methods

### 5.1 Numerical simulation

Simulations of the LMA-PCF were carried out using the MODE Solutions software from Lumerical Solutions, Inc. The pitch and hole diameters of LMA-PCF (LMA-25, NKT Photonics, Inc.) are 16.4 μm and 4 μm, respectively. The material of the photonic crystal fiber used is pure silica glass. Simulations of the metalens' intensity distributions were performed using Computer Simulation Technology Microwave Studio. For the design of unit elements, a unit cell boundary condition is employed for the simulation of transmission spectra in an array configuration (see Supplementary Table S1 for details of the structural parameters of the unit elements for two PCF metalenses). For simplicity, cylindrical lenses are simulated to numerically predicate the focal length of designed metalenses, where a perfectly matched layer and periodic boundary conditions were employed in the $x$ and $y$ directions, respectively. The simulated dimension of the metalens in the $x$ direction is set to 25 μm, which is the actual core diameter of the LMA-PCF. The permittivity function of silica is modeled with the standard Sellmeier expansion [41]. The complex frequency-dependent dielectric function of gold in the NIR regime is described by the Lorentz-Drude model with a damping constant of 0.07 eV and a plasma frequency of 8.997 eV.

## 6 Supplementary material

The supplementary material is available online on the journal's website or from the authors.

**Acknowledgments:** The authors acknowledge the support from the Robert A. Welch Foundation (Funder Id: http://dx.doi.org/10.13039/100000928, award number: AA-1956-20180324), the Young Investigator Development Program, and the Office of Vice Provost for Research at Baylor University. The authors acknowledge the Center for Microscopy and Imaging (CMI) at Baylor University for technical support during the microscopy and image analysis.